\documentclass[aps,11pt, superscriptaddress, showpacs,showkeys,floatfix]{revtex4}
\usepackage{graphicx}
\usepackage{colordvi}

	\newcounter{comentario}
\begin{document}

\title{A canonical transformation and the tunneling probability for the 
birth of an asymptotically DeSitter universe with dust.}

\author{E. V. Corr\^{e}a Silva\footnote{E-mail: evasquez@uerj.br}}

\author{G. A. Monerat\footnote{E-mail: monerat@uerj.br}}
 
\author{G. Oliveira-Neto\footnote{E-mail: gilneto@uerj.br}}

\author{C. Neves\footnote{E-mail: clifford@fat.uerj.br}}

\affiliation{Departamento de Matem\'{a}tica e Computa\c{c}\~{a}o, \\
Faculdade de Tecnologia, \\ 
Universidade do Estado do Rio de Janeiro,\\
Rodovia Presidente Dutra, Km 298, P\'{o}lo
Industrial,\\
CEP 27537-000, Resende-RJ, Brazil.}

\author{L. G. Ferreira Filho\footnote{E-mail: gonzaga@fat.uerj.br}}

\affiliation{Departamento de Mec\^{a}nica e Energia, 
Faculdade de Tecnologia, \\ 
Universidade do Estado do Rio de Janeiro,\\
Rodovia Presidente Dutra, Km 298, P\'{o}lo
Industrial,\\
CEP 27537-000, Resende-RJ, Brazil.}

\date{\today}

\begin{abstract}
In the present work, we study the quantum cosmology description of
closed Friedmann-Robertson-Walker models in the presence of a positive 
cosmological constant and a generic perfect fluid. We work in the
Schutz's variational formalism. If one uses the scale factor and its canonically 
conjugated momentum as the phase space variables that describe the geometrical 
sector of these models, one obtains Wheeler-DeWitt equations with operator ordering 
ambiguities. In order to avoid those ambiguities and simplify the quantum 
treatment of the models, we introduce new phase space variables. We explicitly 
demonstrate that the transformation leading from the old set of variables to the 
new one is canonical. In order to show that the above canonical transformations 
simplify the quantum treatment of those models, we consider a particular model 
where the perfect fluid is dust. We solve the Wheeler-DeWitt equation numerically 
using the Crank-Nicholson scheme and determine the time evolution of the initial 
wave function. Finally, we compare the results for the present model with the ones for 
another model where the only difference is the presence of a radiative perfect fluid, 
instead of dust.
\end{abstract}



\maketitle

\section{Introduction}
\label{sec:intro}

Since the pioneering work in quantum cosmology due to DeWitt \cite{dewitt},
many physicists have worked in this theory. The main motivation behind 
quantum cosmology is a consistent explanation for the origin of our Universe.
So far, the most appealing explanation is the spontaneous {\it creation from 
nothing} \cite{grishchuk, vilenkin, hawking, linde, rubakov, vilenkin1}.
In that picture for the origin of the Universe, the Universe is a quantum
mechanical system with zero size. There is a potential barrier that the 
Universe may tunnel with a well-defined, non-zero probability. If the
Universe actually tunnels, it emerges to the right of the barrier with a
definite size. The application of the {\it creation from nothing} idea in 
minisuperspace models has led to several important results. The wave-function 
of the Universe satisfies the Wheeler-DeWitt equation \cite{wheeler, dewitt}. 
Therefore, one needs to specify boundary conditions in order to solve that 
equation and find a unique and well-defined wave function. The motivation to 
obtain a wave-function that represents the {\it creation from nothing} has led 
to the introduction of at least three proposals for the boundary conditions for 
the wave-function of the Universe \cite{vilenkin1}. The inflationary period of 
the Universe appears very naturally from the {\it creation from nothing} idea. 
That is the case because most of the minisuperspace models considered so far
have a potential that decreases, without a limit, to the right of the barrier.
It gives rise to a period of unbounded expansion which is interpreted as the 
inflationary period of the Universe \cite{vilenkin1}. Also, it was shown by
several authors that an open inflationary universe may be created from nothing,
in theories of a single scalar field for generic potentials \cite{hawking1,
linde1, bousso}. Another important issue is the particle content in the
Universe originated during the {\it creation from nothing} process \cite{rubakov,
rubakov1, vilenkin2}.

In the present work, we study the quantum cosmology description of
Friedmann-Robertson-Walker (FRW) models in the presence of a positive cosmological 
constant ($\Lambda$) and a generic perfect fluid. The spatial sections of the
models have positive curvature. The perfect fluid has an equation of state of the 
form $p = \alpha w$, where $p$ and $w$ are, respectively, the fluid pressure and 
density, and $\alpha$ is a constant in the range $-1 \leq \alpha < 1$. In the present 
quantum cosmology description, the perfect fluid is treated by means of the variational 
formalism developed by Schutz \cite{schutz}. The treatment of a simpler model which 
differs from ours by the lack of a cosmological constant was first done in \cite{germano1}.
The Schutz's variational formalism, applied to the present model, leads to a
superhamiltonian which, after the canonical quantization, produces a Wheeler-DeWitt 
equation with operator ordering ambiguities. More precisely, the canonically conjugated 
momentum to the scale factor ($p_a$) appears in the superhamiltonian coupled to a function 
of the scale factor ($a$). In the present work we propose a general canonical transformation
in order to eliminate these operator ordering ambiguities. As we shall see, the resulting
Wheeler-DeWitt equation in the new variables will be much simpler than the initial one. In fact,
there is already in the literature a particular example of such canonical transformation 
\cite{lemos1}. There, the authors considered a flat FRW geometry coupled to a dust perfect 
fluid ($\alpha = 0$) and a cosmological constant. After the application of Schutz's
variational formalism and the particular canonical transformation, the Wheeler-DeWitt
equation in the new variables was reduced to a Schr\"{o}dinger equation for a harmonic
oscillator when $\Lambda < 0$ and for an inverted harmonic oscillator when $\Lambda > 0$.
Although the authors of \cite{lemos1} state that the transformation they used is canonical, 
they did not prove it. Here, we prove that the transformation leading
from the scale factor and its canonically conjugated momentum to a new set of canonically
conjugated variables, such that in the new variables the resulting Wheeler-DeWitt equation is
free from operator ordering ambiguities, is indeed canonical. 
In order to prove that the transformation is canonical we shall 
demonstrate that the Dirac and Poisson brackets of the new canonical variables are the same.
It will be done following the symplectic framework \cite{JBW}. Then, we shall give a further 
example, than the one given in Ref. \cite{lemos1}, on how these transformations simplify the 
Wheeler-DeWitt equation of these models. We shall consider a closed FRW model coupled to a
dust perfect fluid and a positive cosmological constant. Finally, we compare the results for 
the present model with the ones for another model already treated in the literature \cite{gil1} 
where the only difference is the presence of a radiative perfect fluid, instead of 
dust.

In the next Section, 
we write the superhamiltonian constraint in terms of the scale factor, the
variable associated to the perfect fluid and their conjugated momenta. Then,
the superhamiltonian term which will lead to operator ordering ambiguities, after
the quantization, will be easily seen. After that, we introduce the
transformations leading to new pairs of canonical variables such that the new quantum
theory is free from operator ordering ambiguities. Next, we demonstrate that the
Dirac brackets of the new canonical variables are equal to their corresponding
Poisson brackets, which confirms that the transformations, leading from the old 
variables to the new ones, are canonical. This will be done following the symplectic 
framework \cite{JBW}. In Section \ref{sec:quantum}, we give an explicit example on how 
the canonical transformation simplifies the treatment of the model at the quantum level. 
We consider a specific example where the perfect fluid is dust ($\alpha = 0$). We canonically 
quantize the model, written in terms of the new pair of canonically conjugated variables, 
and obtain the corresponding Wheeler-DeWitt equation. We solve it, numerically.
For particular values of the dust energy and the cosmological constant, we plot 
the square modulus of the wave-function of the universe as a function of the scale factor. 
The tunneling process can be readily seen from the results. Then, we evaluate the tunneling 
probability (TP) and its dependence on the dust energy. We obtain that the TP increases 
with the dust energy for a fixed cosmological constant. Therefore, it is more probable that 
the classical evolution should start with the greatest possible value for the dust 
energy. Finally, we compare the tunneling probability obtained here with the one obtained 
for a model where the perfect fluid is radiative ($\alpha = 1/3$) \cite{gil1}. We find that 
the tunneling probability of the universe with dust is greater than the universe with 
radiation. Finally, in Section \ref{sec:conclusions} we summarize the main points and results 
of our paper.

\section{The classical model and the canonical transformation}
\label{sec:classical}

The closed Friedmann-Robertson-Walker cosmological models are characterized by the
scale factor $a(t)$ and have the following line element,

\begin{equation}  
\label{1}
ds^2 = - N^2(t) dt^2 + a^2(t)\left( \frac{dr^2}{1 - r^2} + r^2 d\Omega^2
\right)\, ,
\end{equation}
where $d\Omega^2$ is the line element of the two-dimensional sphere with
unitary radius, $N(t)$ is the lapse function and we are using the natural
unit system, where $\hbar=c=G=1$. The matter content of the model is
represented by a perfect fluid with four-velocity $U^\mu = \delta^{\mu}_0$
in the comoving coordinate system used and a positive cosmological
constant ($\Lambda$). The total energy-momentum tensor is given by,

\begin{equation}
T_{\mu,\, \nu} = (w+p)U_{\mu}U_{\nu} - p g_{\mu,\, \nu} - \Lambda
g_{\mu,\, \nu}\, ,  
\label{2}
\end{equation}
where $w$ and $p$ are, respectively, the fluid density and pressure.
Here, we assume that $p = \alpha w$, where $-1 \leq \alpha < 1$, which 
is the equation of state for a perfect fluid. 

Einstein's equations for the metric (\ref{1}) and the energy momentum 
tensor (\ref{2}) are equivalent to the Hamilton equations generated by 
a total Hamiltonian. In order to get this Hamiltonian, the symplectic 
method will be applied in a seminal Lagrangian that describes the 
dynamics of the system,

\begin{equation}
\label{b0000}
{\cal L}=-\frac{3a{\dot a}^{(2-3\alpha)}}{N} + 3Na^{3\alpha} - \Lambda N a^{(2+3\alpha)}- \left[\frac{3a^{3/4}}{4N^{3/4}}\left(\dot\epsilon+\theta\dot B\right)\right]^4 e^{-3B},
\end{equation}
which is an extended formulation of Schutz's Lagrangian, since the 
$\alpha$ parameter was introduced into the model. Calculating the canonical 
momenta, namely,

\begin{eqnarray}
\label{b0010}
P_a&=& -\frac{6a^{(2-3\alpha)}\dot a}{N^{3/4}},\\
P_\epsilon &=&- \frac{3a^{3/4}}{N^{3/4}}\left[\frac{3a^{3/4}}{4N^{3/4}}\left(\dot\epsilon
+\theta\dot B\right)\right]^3 e^{-3B},\\
P_B &=&- \frac{3a^{3/4}\theta}{N^{3/4}}\left[\frac{3a^{3/4}}{4N^{3/4}}\left(\dot\epsilon
+\theta\dot B\right)\right]^3 e^{-3B},
\end{eqnarray}
the Lagrangian density, eq.(\ref{b0000}), can be rewritten in a first-order form, given by

\begin{equation}
\label{b0020}
{\cal L}=P_a\dot a + P_\epsilon\dot\epsilon + P_B\dot B +
\frac{N{P_a}^2}{12a^{(2-3\alpha)}} + 3Na^{3\alpha} - 
\Lambda N a^{(2+3\alpha)}- \left[\frac{3a^{3/4}}{4N^{3/4}}\left(\dot\epsilon
+\theta\dot B\right)\right]^4 e^{-3B}.
\end{equation}
Introducing the following transformation

\begin{eqnarray}
\label{b0030}
P_T &=& \frac{P_\varepsilon^{4/3} e^B}{3^{1/3}},\nonumber\\
\dot\varepsilon &=& \frac{P_\varepsilon^{1/3}  e^B}{2 . 3^{1/3}}\dot{T},\\
\dot{B} &=& \frac{\dot\varepsilon}{\theta},\nonumber
\end{eqnarray}
the Lagrangian density, eq.(\ref{b0020}), becomes

\begin{eqnarray}
\label{b0040}
{\cal L}^{(0)}
&=& P_a\dot a + P_T\dot T +\frac{N{P_a}^2}{12a^{(2-3\alpha)}} 
+ 3Na^{3\alpha} - \Lambda N a^{(2+3\alpha)}- N\frac{P_T}{a},\nonumber\\
\mbox{}&=& P_a\dot a + P_T\dot T - V,
\end{eqnarray}
where $V=N\Omega$ is the symplectic potential, with

\begin{equation}
\label{b0050}
\Omega = -\frac{{P_a}^2}{12a^{(2-3\alpha)}} - 3a^{3\alpha} + 
\Lambda a^{(2+3\alpha)}+ \frac{P_T}{a}.
\end{equation}

Now, we are ready to apply the symplectic method. The symplectic 
variables are 
\mbox{$\xi^{(0)}_i=(a,P_a,T,P_T,N)$} 
and the one-form momenta 
$(A_i)$ are given by

\begin{eqnarray}
\label{b0070}
A_a &=& P_a,\nonumber\\
A_{P_a} &=& 0,\nonumber\\
A_{T} &=& P_T,\\
A_{P_T} &=& 0,\nonumber\\
A_{N} &=& 0.\nonumber
\end{eqnarray}
The corresponding symplectic matrix, 
defined as

\begin{equation}
\label{b0060}
f_{ij}=\frac{\partial A_j}{\partial\xi_i}-\frac{\partial A_i}{\partial\xi_j},
\end{equation}
reads

\begin{equation}
\label{b0080}
F^{(0)}=
\left(\begin{array}{ccccc}
0 & -1 & 0 & 0 & 0 \\1 & 0 & 0 & 0 & 0 \\0 & 0 & 0 & -1 & 0 
\\0 & 0 & 1 & 0 & 0 \\0 & 0 & 0 & 0 & 0
\end{array}\right).
\end{equation}
This matrix is singular, so it has a zero-mode,

\begin{equation}
\label{b0090}
\nu=\left(\begin{array}{ccccc}
0 & 0 & 0 & 0 & 1
\end{array}\right).
\end{equation}
Following the symplectic method, if one contracts the 
zero-mode with the symplectic potential gradient one should 
find zero. Otherwise, a constraint is obtained. Contracting 
$\nu$ eq.(\ref{b0090}) with the symplectic potential 
gradient, we get

\begin{equation}
\label{b0100}
\nu_i\frac{\partial V}{\partial x_i}=\Omega,
\end{equation}
which is a constraint. This constraint will be introduced 
into the first-order Lagrangian, eq.(\ref{b0040}), through 
a Lagrange multiplier, $\beta$,

\begin{equation}
\label{b0110}
{\cal L}^{(1)} 
= P_a\dot a + P_T\dot T + \Omega\dot\beta- V.
\end{equation}
Now, the 
symplectic variables are 
$\xi^{(1)}_i=(a,P_a,T,P_T,N,\beta)$
and the respective symplectic matrix is given by

\begin{equation}
\label{b0120}
F^{(1)}
=\left(\begin{array}{cccccc}
0 & -1 & 0 & 0 & 0 & \frac{\partial\Omega}{\partial a}\\
1 & 0 & 0 & 0 & 0 & \frac{\partial\Omega}{\partial P_a}\\
0 & 0 & 0 & -1 & 0 & 0\\0 & 0 & 1 & 0 & 0 & \frac{\partial\Omega}{\partial P_T}\\
0 & 0 & 0 & 0 & 0 & 0\\ -\frac{\partial\Omega}{\partial a} 
& -\frac{\partial\Omega}{\partial P_a} & 0 & 
-\frac{\partial\Omega}{\partial P_T} & 0 & 0\\
\end{array}\right).
\end{equation}
This matrix is singular and, therefore, has a zero-mode,

\begin{equation}
\label{b0130}
\kappa=\left(\begin{array}{cccccc}
-\frac{\partial\Omega}{\partial P_a}& \frac{\partial\Omega}{\partial a} 
& -\frac{\partial\Omega}{\partial P_T} & 0 & 1 & 1
\end{array}\right).
\end{equation}
Contracting this zero-mode with the symplectic potential 
gradient, we get

\begin{equation}
\label{b0140}
\kappa_i\frac{\partial V}{\partial x_i}=\Omega.
\end{equation}
This constraint was obtained before and, following the 
symplectic method, the system has a gauge symmetry. 
In order to quantize the system, the symmetry must be 
fixed. This is done in the symplectic process introducing 
the gauge fixing term into the original first-order 
Lagrangian, eq.(\ref{b0040}), through a Lagrange multiplier, $\eta$,

\begin{equation}
\label{b0150}
{\cal L}^{(2)} = 
P_a\dot a + P_T\dot T + \Sigma\dot\eta - V,
\end{equation}
where the gauge fixing term is

\begin{equation}
\label{b0160}
\Sigma=N - a.
\end{equation}
The symplectic variables 
are, now, $\xi^{(2)}_i=(a,P_a,T,P_T,N,\eta)$ 
and the symplectic matrix is

\begin{equation}
\label{b0170}
F^{(2)}
=\left(\begin{array}{cccccc}
0 & -1 & 0 & 0 & 0 & -1\\1 & 0 & 0 & 0 & 0 & 0\\
0 & 0 & 0 & -1 & 0 & 0\\0 & 0 & 1 & 0 & 0 & 0 \\
0 & 0 & 0 & 0 & 0 & 1\\1 & 0 & 0 & 0 & -1 & 0.
\end{array}\right).
\end{equation}
This is a nonsingular matrix and, according to the 
symplectic process, its inverse allows us to get the 
Dirac brackets; the only non-vanishing brackets are

\begin{equation}
\label{b00180}
\left\{a,P_a\right\}=\left\{T,P_T\right\}=1.
\end{equation}
The symplectic potential 
is identified as being the Hamiltonian, so

\begin{equation}
\label{b00190}
{\cal H} =-\frac{{P_a}^2}{12a^{(1-3\alpha)}} - 
3a^{1+3\alpha} + \Lambda a^{(3+3\alpha)}+ {P_T}.
\end{equation}
where $P_{a}$ and $P_{T}$ are the momenta canonically 
conjugated to $a$ and $T$, the latter being the canonical 
variable associated to the fluid \cite {germano1}. This 
is the total Hamiltonian that generates Hamilton's equations  
equivalent to Einstein's equations for the metric (\ref{1}) and 
the energy-momentum tensor (\ref{2}). Since a quantization 
process will be applied, the first term in the total Hamiltonian  
(\ref{b00190}) poses an operator ordering ambiguity. In order to 
solve this problem, the following transformation will be applied,

\begin{equation}
\label{b00200}
a = \left(\frac {3(1-\alpha)x}{2}\right)^{2/3(1-\alpha)}.
\end{equation}
Due to this, the Lagrangian density, eq.(\ref{b0000}), becomes

\begin{eqnarray}
\label{b0210}
{\cal L}^{'(0)}
&=&-\frac{3}{N}\left(\frac{3(1-\alpha)x}{2}\right)^{2/3(1-\alpha)}{\dot x}^2 
+ 3N \left(\frac{3(1-\alpha)x}{2}\right)^{2\alpha/(1-\alpha)}  - \Lambda N \left(\frac{3(1-\alpha)x}{2}\right)^{2(2+3\alpha)/3(1-\alpha)}\nonumber\\
\mbox{}&-& \left[\frac{\left(3(1-\alpha)x\right)^{1/2(1-\alpha)}}{2^{2(4-3\alpha)/3(1-
\alpha)}N^{3/4}}\left(\dot\epsilon+\theta\dot B\right)\right]^4 e^{-3B}.
\end{eqnarray}

Using the transformation given in eq.(\ref{b0030}) and with the aid of the momenta $P_x$, 
$P_\epsilon$, $P_B$ computed from ${\cal L}^{'(0)}$ (\ref{b0210}), the first-order 
Lagrangian will become

\begin{equation}
\label{b0220}
{\cal L}^{'(1)} = 
P_x\dot x + P_T\dot T - V,
\end{equation}
where the symplectic potential is $V=N\Omega$, with

\begin{eqnarray}
\label{b0230}
\Omega &=& -\frac{{P_x}^2}{12\left(\frac{3(1-\alpha)x}{2}\right)^{2/3(1-\alpha)}} - 3 \left(\frac{3(1-\alpha)x}{2}\right)^{2\alpha/(1-\alpha)} \nonumber\\
&+& \Lambda \left(\frac{3(1-\alpha)x}{2}\right)^{2(2+3\alpha)/3(1-\alpha)} + \frac{P_T}{\left(\frac{3(1-\alpha)x}{2}\right)^{2/3(1-\alpha)}}.
\end{eqnarray}
After that, the symplectic method could be applied. The 
symplectic variables 
are now \mbox{$\xi^{'(1)}_i=(x,P_x,T,P_T,N)$}
with the corresponding one-form momenta,

\begin{eqnarray}
\label{b0070a}
A_x &=& P_x,\nonumber\\
A_{P_x} &=& 0,\nonumber\\
A_{T} &=& P_T,\\
A_{P_T} &=& 0,\nonumber\\
A_{N} &=& 0.\nonumber\\
\end{eqnarray}
Computing the corresponding symplectic matrix, using 
eq.(\ref{b0060}), one finds that it has the following
zero-mode,


\begin{equation}
\label{b0090a}
\nu=\left(\begin{array}{ccccc}
0 & 0 & 0 & 0 & 1
\end{array}\right).
\end{equation}
Contracting this zero-mode with the 
symplectic potential gradient, a 
constraint is obtained, namely,

\begin{equation}
\label{b0100a}
\nu_i\frac{\partial V}{\partial x_i}=\Omega
\end{equation}
This constraint will be introduced into the first-order 
Lagrangian (\ref{b0220}) through a Lagrange multiplier, 
$\beta$, yielding

\begin{equation}
\label{b0110a}
{\cal L}^{'(2)} = 
P_a\dot a + P_T\dot T + \Omega\dot\beta- V.
\end{equation}
The symplectic variables 
are, now, $\xi^{'(2)}_i=(a,P_a,T,P_T,N,\beta)$ 
and the respective symplectic matrix is singular
with the following zero-mode,

\begin{equation}
\label{b0130a}
\kappa=\left(\begin{array}{cccccc}
-\frac{\partial\Omega}{\partial P_a}& \frac{\partial\Omega}{\partial a} 
& -\frac{\partial\Omega}{\partial P_T} & 0 & 1 & 1
\end{array}\right).
\end{equation}
Contracting this zero-mode with the gradient of the 
symplectic potential, we get

\begin{equation}
\label{b0140a}
\kappa_i\frac{\partial V}{\partial x_i}=\Omega.
\end{equation}
This constraint was obtained before and, in agreement 
with the symplectic method, the system has a gauge 
symmetry. This symmetry must be fixed, to this end a 
gauge fixing term $(\Sigma)$ is introduced into the 
original first-order Lagrangian (\ref{b0220}),

\begin{equation}
\label{b0240}
\overline{\cal L}
 = P_x\dot x + P_T\dot T + \Sigma\dot\eta - V,
\end{equation}
with

\begin{equation}
\label{b0250}
\Sigma=N - \left(\frac {3(1-\alpha)x}{2}\right)^{2/3(1-\alpha)}.
\end{equation}
Now, the symplectic variables are 
$\overline{\xi}_i=(x,P_x,T,P_T,N,\eta)$ 
and the corresponding symplectic matrix is nonsingular.
From the inverse of the symplectic matrix 
the non-vanishing Dirac brackets
\begin{equation}
\label{b00260}
\left\{x,P_x\right\}=\left\{T,P_T\right\}=1
\end{equation}
are obtained; the remaining brackets are all null.
The symplectic potential 
is identified as being the Hamiltonian, so

\begin{equation}
\label{b00270}
{\cal H} = N\Omega = -\frac{{P_x}^2}{12} - V_{eff.}+ {P_T},
\end{equation}
with

\begin{equation}
\label{b00280}
V_{\mbox{\it eff.}}
= 3 \left(\frac{3(1-\alpha)x}{2}\right)^{(2+6\alpha)/3(1-\alpha)}
- \Lambda \left(\frac{3(1-\alpha)x}{2}\right)^{2(1+\alpha)/(1-\alpha)}.
\end{equation}
Note that the Hamiltonian above is a Schr\"{o}dinger-like
equation with no operator ordering ambiguities. Further, the 
Dirac brackets are equal to the Poisson brackets, allowing us
to conclude that the variable transformation, eq.(\ref{b00200}), 
is a canonical transformation.

\section{The quantization of a particular model}
\label{sec:quantum}

In order to show the usefulness of transformation (\ref{b00200}), let us 
consider a particular example. Consider a model with closed FRW geometry 
coupled to a dust perfect fluid ($\alpha=0$) and a positive cosmological 
constant ($\Lambda$). The total Hamiltonian, eq.(\ref{b00270}), reduces to

\begin{equation}
\label{4a}
{\cal H} =-\frac{{P_x}^2}{12} - 3 \left(\frac{3x}{2}\right)^{2/3}
+ \Lambda \left(\frac{3x}{2}\right)^{2}+ {P_T}.
\end{equation}

The model will be quantized following the Dirac formalism for
constrained systems \cite{dirac}. First we introduce a wave-function 
which depends on the canonical variables $\hat{x}$ and $\hat{T}$,

\begin{equation}  
\label{7}
\Psi\, =\, \Psi(\hat{x} ,\hat{T} )\, .
\end{equation}
Then, we impose the appropriate commutators between the operators $\hat{x}$
and $\hat{T}$ and their respective conjugate momenta $\hat{P}_x$ and $\hat{P}_T$.
Working in the Schr\"{o}dinger picture, the operators $\hat{x}$ and $\hat{T}$
are simply multiplication operators, while their conjugate momenta are
represented by the differential operators 
\begin{equation}
P_{x}\rightarrow -i\frac{\partial}{\partial x}\hspace{0.2cm},\hspace{0.2cm} 
\hspace{0.2cm}
P_{T}\rightarrow -i\frac{\partial}{\partial T}\hspace{0.2cm}.
\label{8}
\end{equation}

Finally, we demand that the operator corresponding to $\mathcal{H}$ 
annihilate the wave-function $\Psi$, which leads to Wheeler-DeWitt 
equation 
\begin{equation}
\bigg(\frac{1}{12}\frac{{\partial}^2}{\partial x^2} - 
3\left({3x\over 2}\right)^{2/3} + \Lambda \left({3x\over 2}\right)^{2}
\bigg)\Psi(x,\tau) = -i \, \frac{\partial}{\partial \tau}\Psi(x,\tau),
\label{9}
\end{equation}
where the new variable $\tau= -T$ has been introduced.

The operator $\hat{\mathcal{H}}$ is self-adjoint \cite{lemos} with respect
to the internal product,

\begin{equation}
(\Psi ,\Phi ) = \int_0^{\infty} dx\, \,\Psi(x,\tau)^*\, \Phi (x,\tau)\, ,
\label{10}
\end{equation}
if the wave functions are restricted to the set of those satisfying either 
$\Psi (0,\tau )=0$ or $\Psi^{\prime}(0, \tau)=0$, where the prime $\prime$
means the partial derivative with respect to $x$. Here, we consider wave 
functions satisfying the former type of boundary condition and we also 
demand that they vanish when $x \rightarrow \infty$.

The Wheeler-DeWitt equation (\ref{9}) is a Schr\"{o}dinger equation for
a potential with a barrier. We solve it numerically using a 
finite-difference procedure based on the Crank-Nicholson
method \cite{crank}, \cite{numericalrecipes} and implemented in the 
program GNU-OCTAVE. Our choice of the Crank-Nicholson 
method is based on its recognized stability.
The norm conservation is commonly used as a criterion to evaluate the
reliability of the numerical calculations of the time evolution of
wave functions. In References \cite{Iitaka} and \cite{teukolsky}, this 
criterion is used to show analytically that the Crank-Nicholson method is 
unconditionally stable. Here, in order to evaluate the reliability of our 
algorithm, we have numerically calculated the norm of the wave packet for 
different times. The results thus obtained show that the norm is preserved.
  
In fact, numerically one can only treat
the {\it tunneling from something} process, where one gives a initial wave
function with a given mean energy, very concentrated in a region next to 
$x=0$. That initial condition fixes an energy for the dust and the 
initial region where $x$ may take values. Our choice for the initial wave
function is the following normalized gaussian,

\begin{equation}
\label{11}
\Psi(x,0) = \left({8192 E^3\over \pi}\right)^{1/4} x 
e^{(-4 E x^2)}\, ,
\end{equation}
where $E$ is the dust energy. The wave-function $\Psi(x,0)$ is normalized by 
demanding that the integral of $|\Psi(x,0)|^2$ from $0$ to $\infty$ be equal 
to one and its mean energy be $E$. After one gives the initial wave function,
one leaves it evolve following the appropriate Schr\"{o}dinger equation
until it reaches infinity in the $x$ direction. Numerically, one has to 
fix the infinity at a finite value, let us call it $x_{maxd}$. The general 
behavior of the solutions, when $E$ is smaller than the maximum value of the 
potential barrier, is an everywhere well-defined, finite, normalized wave 
packet. Even in the limit of a vanishing scale factor.
For small values of $x$ the wave packet has
great amplitudes and oscillates rapidly due to the interaction between
the incident and reflected components. The transmitted component is an
oscillatory wave packet that moves to the right and has a decreasing 
amplitude which goes to zero in the limit when $x$ goes to infinity.
As an example, we solve eq. ({\ref{9}) with $\Lambda=0.01$. For this
choice of $\Lambda$ the potential barrier has its maximum value equal to
$20$. In order to see the tunneling process, we choose $E=19$ for the
initial wave function eq. (\ref{11}). For that energy, we compute the points
where it meets the potential barrier, the left ($x_{ltp}$) and right
($x_{rtp}$) turning points. They are, $x_{ltp}=15.4085707$ and $x_{rtp}=26.933766$.
In the present case we fix $x_{maxd}=83.23582897$, as the infinity in the $x$ 
direction. In figure \ref{fig1}, we show $|\Psi(x,t_{maxd})|^2$ for the values of
$\Lambda$ and $E$, given above, at the moment $t_{maxd}=21.5$ when $\Psi$ reaches
infinity. For more data on this particular case see Table \ref{tableenergy} in 
the appendix. It is important to mention that the particular choice of
numerical values for $\Lambda$ and $E$, above and in the following examples, 
were made simply for a better visualization of the different properties of the 
system.

\begin{figure}[h!]
\includegraphics[width=7cm,height=9cm,angle=-90]{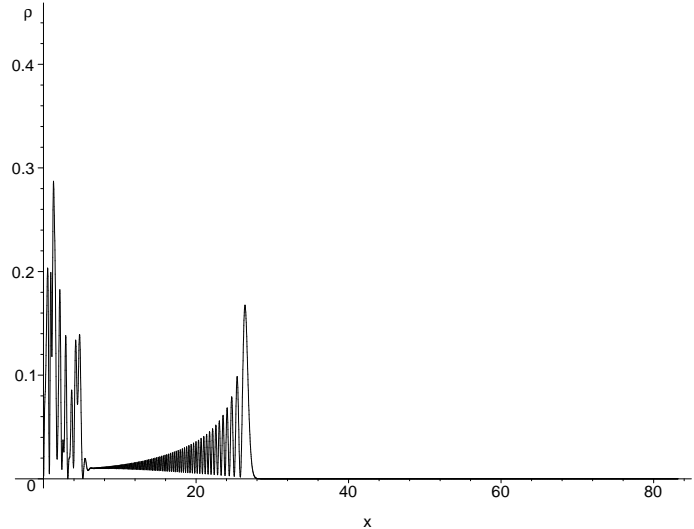}
\caption{{\protect\footnotesize {$|\Psi(x,t_{maxd})|^2 \equiv \rho$, for 
$\Lambda=0.01$, $E=19$ at the moment $t_{maxd}=21.5$ when $\Psi$ reaches 
infinity, located at $x=83.23582897$.}}}
\label{fig1}
\end{figure}

Now, we compute the tunneling probability as the probability to find the 
scale factor of the universe to the right of the potential barrier.
In the present situation, this definition is given by the following 
expression \cite{gil1},

\begin{equation}
\label{12}
TP_{int} = {\int_{x_{rtp}}^{\infty} |\Psi(x,t_{max})|^2 dx \over 
\int_{0}^{\infty} |\Psi(x,t_{max})|^2 dx} \, ,
\end{equation}
where, as we have mentioned above, the $\infty$ limit must be replaced by 
some suitably large value of $x$, for the sake of numerical calculations.

Since, by normalization, the denominator of Eq. (\ref{12}) is equal to the
identity, $TP_{int}$ is effectively given by the numerator of Eq. (\ref{12}). 
We consider, here, the dependence of TP on the energy $E$. Therefore, we 
compute $TP_{int}$ for many different values of $E$ for a fixed $\Lambda$.
For all cases, we consider the situation where $E$ is smaller than the
maximum value of the potential barrier. From that numerical study we 
conclude that the tunneling probability grows with $E$ for a fixed $\Lambda$.
As an example, we consider $20$ values of the dust energy for a fixed 
$\Lambda=0.01$. For this choice of $\Lambda$ the potential barrier has its 
maximum value equal to $20$. In order to study the tunneling process, we fix 
the mean energies of the various $\Psi(x,0)$'s in Eq. (\ref{11}) to be smaller 
than that value. Table \ref{tableenergy} in the appendix shows} the different 
energy values $E$,  $TP_{int}$, $x_{ltp}$ and $x_{rtp}$ for each energy. 


Since $TP$ grows with $E$ it is more likely for the universe, described by 
the present model, to nucleate with the highest possible dust energy.
Therefore, it is more probable that the classical evolution should start 
with the greatest possible value for the dust energy.

\begin{figure}[h!]
\includegraphics[width=7cm,height=9cm,angle=-90]{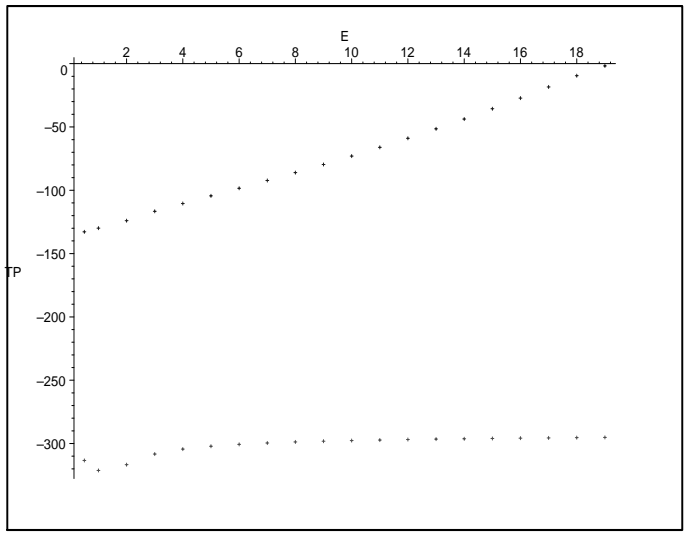}
\caption{{\protect\footnotesize {
$\log TP_{int}$ for different energies ($E$) for a fixed $\Lambda=0.01$,
for both radiative and dust models.}}}
\label{fig3}
\end{figure}

Now, we may compare the above $TP'$s for dust with the $TP'$s for a radiative
perfect fluid ($\alpha = 1/3$). The $TP'$s for a closed FRW model coupled to 
a radiative perfect fluid and a positive cosmological constant, described in 
Schutz's formalism, were first computed in Ref. \cite{gil1}. In order to
compare the $TP'$s between the two models we must fix values for $\Lambda$
and $E$. Then, we must fix values for the infinity ($x_{max}$) in both models 
so that they be comparable. Here, we consider the distance from the point 
$x_{0r}$ where the potential vanishes ($x_{0r}\neq 0$) to $x_{maxr}$, for the 
radiative model, as the reference. We compute the ratio $x_{maxr}/x_{0r}$. Let 
us call it $\Delta_r$. Then, we fix the $x_{maxd}$, for the dust model, by 
multiplying $x_{0d}$, for this model, by $\Delta_r$. Finally, for fixed values
of $\Lambda$, $E$ and $x_{max}$, we compute the $TP'$s for both radiative and
dust models, with the aid of eq. (\ref{12}). We repeat these calculations for
several different energy values $E$ and conclude that the $TP$ for the dust is 
greater than the one for the radiation, for the same energy. As an example, we 
consider the same $20$ values of the dust energy, given in table \ref{tableenergy}, 
and repeat the calculations for the radiative model with fixed $\Lambda=0.01$. For 
this choice of $\Lambda$ the potential barrier, in the radiative model, has its 
maximum value equal to $225$ and $x_{0r} = 17.32050808$. We choose $x_{maxr} = 30$, 
in the radiative model. Therefore, $\Delta_r = (30/17.32050808) = 1.732050807$ 
and since $x_{0d} = 48.05622828$, for the dust model, $x_{maxd} = 83.23582897$.
In table \ref{tableTPradiative}, in the appendix, we can see the different energy 
values $E$, $TP_{int}$, $x_{ltp}$ and $x_{rtp}$ for each energy, in the radiative
model. 
Figure \ref{fig3} shows the tunneling probability as a function
of $E$ in logarithmic scale, for both models.

\section{Conclusions.}
\label{sec:conclusions}

In the present work, we studied the quantum cosmology description of
closed Friedmann-Robertson-Walker (FRW) models in the presence of a positive 
cosmological constant ($\Lambda$) and a generic perfect fluid. We worked in 
the Schutz's variational formalism. If one uses the scale factor and its 
canonically conjugated momentum as the phase space variables that describe
the geometrical sector of these models, one obtains Wheeler-DeWitt equations with 
operator ordering ambiguities. In order to avoid such ambiguities and simplify
the quantum treatment of the models, we introduced new phase space variables. 
We explicitly demonstrated that the transformation leading from the old set of 
variables to the new one is canonical. In the demonstration, carried out in the 
symplectic framework, we showed that the Dirac and Poisson brackets of the new 
canonical variables are the same. In order to show that the above canonical 
transformations simplify the quantum treatment of those models, we considered a 
particular model. We applied it to the canonical quantization of a closed FRW 
model in the presence of a positive cosmological constant and dust. We used the 
Schutz's variational formalism and the appropriate canonical transformation to
re-write the Wheller-DeWitt equation as a Schr\"{o}dinger equation, in
the new variables, free from operator ordering ambiguities. In the present case, the 
Schr\"{o}dinger equation had a potential barrier. We solved it numerically using 
the Crank-Nicholson scheme and determined the time evolution of the initial wave 
function. Then, using the wave function we computed the tunneling probability (TP) 
for the birth of an asymptotically DeSitter, inflationary universe, as a function 
of the mean energy $E$ of the initial wave function. We observed that the TP grows 
with $E$. Finally, we compared the TP of the present model with the TP of another
model already studied in the literature \cite{gil1}. There, the only difference is 
the presence of a radiative perfect fluid ($\alpha = 1/3$), instead of dust. We 
observed that the $TP$ for the dust is greater than the one for the radiation, for 
the same energy. 

\begin{acknowledgements}
E. V. Corr\^{e}a Silva, G. A. Monerat and G. Oliveira-Neto (Researchers of CNPq, 
Brazil) thank CNPq and FAPERJ for partial financial support. C. Neves (Researcher 
of CNPq, Brazil) and L. G. Ferreira Filho thank CNPq for partial financial support.
We thank the opportunity to use the Laboratory for Advanced Computation (LCA) of 
the Departament of Mathematics and Computation, FAT/UERJ, where part of this work 
was prepared.
\end{acknowledgements}

\appendix

\section{Tables}

\begin{table}[ht!]
{\scriptsize\begin{tabular}{|c|c|c|c|}
\hline E & $TP_{int}$ & $x_{ltp}$ & $x_{rtp}$ \\ \hline 
19.000000 & 0.0162028   & 15.408570 & 26.933766 \\ \hline 
18.000000 & 3.15309e-10 & 13.130081 & 29.397063 \\ \hline 
17.000000 & 3.98196e-19 & 11.418328 & 31.300238 \\ \hline 
16.000000 & 6.5559e-28  & 10.004429 & 32.912664 \\ \hline 
15.000000 & 2.27017e-36 & 8.784433 & 34.338844 \\ \hline 
14.000000 & 1.78018e-44 & 7.705290 & 35.632446 \\ \hline 
13.000000 & 3.07762e-52 & 6.735782 & 36.825393 \\ \hline 
12.000000 & 1.11065e-59 & 5.855886 & 37.938510 \\ \hline 
11.000000 & 7.86966e-67 & 5.052047 & 38.986278 \\ \hline 
10.000000 & 1.02929e-73 & 4.314784 & 39.979254 \\ \hline 
9.000000  & 2.3364e-80  & 3.637385 & 40.925413 \\ \hline 
8.000000 & 8.63572e-87  & 3.015151 & 41.830959 \\ \hline 
7.000000 & 4.84268e-93  & 2.444963 & 42.700829 \\ \hline 
6.000000 & 3.77963e-99  & 1.925061 & 43.539032 \\ \hline 
5.000000 & 3.65203e-105 & 1.454985 & 44.348872 \\ \hline 
4.000000 & 3.65156e-111 & 1.035704 & 45.133115 \\ \hline 
3.000000 & 2.47116e-117 & 0.670037 & 45.894101 \\ \hline 
2.000000 & 9.73233e-125 & 0.363698 & 46.633833 \\ \hline 
1.000000 & 1.2813e-130  & 0.128371 & 47.354041 \\ \hline 
0.500000 & 1.45689e-133 & 0.045367 & 47.707300 \\ \hline 
\end{tabular}
}
\caption{{\protect\footnotesize {The computed values of $TP_{int}$,
$x_{ltp}$ and $x_{rtp}$ for $20$ different values of $E$
when $\Lambda=0.01$, $x_{maxd}=83.23582897$ and $t_{maxd}=21.5$,
for the dust model.}}}
\label{tableenergy}
\end{table}

\begin{table}[ht!]
{\scriptsize\begin{tabular}{|c|c|c|c|}
\hline E & $TP_{int}$ & $x_{ltp}$ & $x_{rtp}$ \\ \hline 
19.000000 & 6.4146e-296  & 2.544209 & 17.132630 \\ \hline 
18.000000 & 4.2118e-296  & 2.474885 & 17.142781 \\ \hline 
17.000000 & 2.68803e-296 &2.403736 & 17.152902 \\ \hline 
16.000000 & 1.65734e-296 &2.330596 & 17.162993 \\ \hline 
15.000000 & 9.79524e-297 &2.255268 & 17.173054 \\ \hline 
14.000000 & 5.49363e-297 &2.177524 & 17.183084 \\ \hline 
13.000000 & 2.8852e-297  &2.097094 & 17.193086 \\ \hline 
12.000000 & 1.39383e-297 &2.013655 & 17.203058 \\ \hline 
11.000000 & 6.04368e-298 &1.926814 & 17.213001 \\ \hline 
10.000000 & 2.27188e-298 &1.836087 & 17.222914 \\ \hline 
9.000000 & 7.03829e-299  &1.740866 & 17.232800 \\ \hline 
8.000000 & 1.664e-299    &1.640366 & 17.242656 \\ \hline 
7.000000 & 2.65575e-300  &1.533548 & 17.252485 \\ \hline 
6.000000 & 2.32358e-301  &1.418983 & 17.262285 \\ \hline 
5.000000 & 7.59041e-303  &1.294616 & 17.272057 \\ \hline 
4.000000 & 4.17488e-305  &1.157287 & 17.281802 \\ \hline 
3.000000 & 5.36776e-309  &1.001676 & 17.291519 \\ \hline 
2.000000 & 1.90162e-317  &0.817407 & 17.301209 \\ \hline 
1.000000 & 5.92879e-322  &0.577672 & 17.310872 \\ \hline 
0.500000 & 4.55125e-314  &0.408362 & 17.315693 \\ \hline \end{tabular}
}
\caption{{\protect\footnotesize {The computed values of $TP_{int}$,
$x_{ltp}$ and $x_{rtp}$ for $20$ different values of $E$
when $\Lambda=0.01$, $x_{maxr}=30.0$ and $t_{maxr}=15.0$, for the 
radiative model.}}}
\label{tableTPradiative}
\end{table}

\end{document}